\documentclass[runningheads]{llncs}
\usepackage[T1]{fontenc}
\usepackage{graphicx}
\usepackage{cite}
\usepackage{amsmath,amssymb,amsfonts}
\usepackage{algorithmic}
\usepackage{graphicx}
\usepackage{textcomp}
\usepackage{amsmath,amssymb,amsfonts}
\usepackage{algorithmic}
\usepackage{graphicx}
\usepackage{textcomp}
\usepackage{xcolor}
\usepackage{tikz}
\usepackage[edges]{forest}
\usepackage{array}
\newcolumntype{C}[1]{>{\centering\arraybackslash}p{#1}}
\newcolumntype{L}[1]{>{\raggedright\arraybackslash}p{#1}}
\usetikzlibrary{arrows.meta,shadows}
\usepackage{enumitem}   
\usepackage{tabularx}
\usepackage{booktabs}   
\usepackage{caption}    
\usepackage{graphicx}
\usetikzlibrary{shapes.geometric, arrows, positioning, chains}
\usepackage{xcolor}
\def\BibTeX{{\rm B\kern-.05em{\sc i\kern-.025em b}\kern-.08em
    T\kern-.1667em\lower.7ex\hbox{E}\kern-.125emX}}
\usepackage{hyperref} 
\usepackage{wheelchart}
\usepackage{pgfplots}

\usepackage{xcolor} 
\begin{document}
\title{Optimizing Peer Grading: A Systematic Literature Review of Reviewer Assignment Strategies and Quantity of Reviewers
}
\author{
  Uchswas Paul\inst{1} \and
  Shail Shah\inst{1} \and
  Sri Vaishnavi Mylavarapu\inst{1} \and
  M.\,Parvez Rashid\inst{2} \and
  Edward Gehringer\inst{1}
}
%
\authorrunning{U. Paul et al.}
\institute{
North Carolina State University, Raleigh, NC, USA\\
  \email{\{upaul, sshah38, smylava, efg\}@ncsu.edu}
  \and
  College of Charleston, Charleston, SC, USA\\
  \email{rashidp@cofc.edu}
}

\maketitle          
\begin{abstract}
Peer assessment has established itself as a critical pedagogical tool in academic settings, offering students timely, high-quality feedback to enhance learning outcomes. However, the efficacy of this approach depends on two factors: (1) the strategic allocation of reviewers and (2) the number of reviews per artifact. This paper presents a systematic literature review of 87 studies (2010--2024) to investigate how reviewer-assignment strategies and the number of reviews per submission impact the accuracy, fairness, and educational value of peer assessment.
We identified four common reviewer-assignment strategies: random assignment, competency-based assignment, social-network-based assignment, and bidding. Drawing from both quantitative data and qualitative insights, we explored the trade-offs involved in each approach. Random assignment, while widely used, often results in inconsistent grading and fairness concerns. Competency-based strategies can address these issues. Meanwhile, social and bidding-based methods have the potential to improve fairness and timeliness---existing empirical evidence is limited.
In terms of review count, assigning three reviews per submission emerges as the most common practice. A range of three to five reviews per student or per submission is frequently cited as a recommended spot that balances grading accuracy, student workload, learning outcomes, and engagement.

\end{abstract}

\section{Introduction}
 Peer grading has attracted growing interest due to its scalability, especially with the rise of MOOCs and the increasing size of classes\cite{Formanek2017}. Beyond reducing the grading burden on instructors~\cite{Price2016}, evaluating classmates' work gives students an understanding of the requirements and alternative approaches to a problem \cite{Chiou2015, Wimshurst2013}. Despite many advantages, the peer grading system is not free of noise and bias~\cite{Namanloo2022, Jimenez-Romero2013, Capuano2016}. The effectiveness of peer grading largely depends on how the process is designed and implemented. 
 
Several strategies have been proposed to address inaccuracies in peer grading. These include weighting grades using a calibration score~\cite{cali}, leveraging reputation~\cite{García-Martínez2019}, or relying on agreement among reviewers~\cite{Walsh2014}. While these methods primarily focus on correcting errors after grades have been assigned, preventive measures also play a vital role in enhancing accuracy, reliability, and learning outcomes. The first step in prevention is how assignments are distributed, which involves assigning appropriate reviewers, deciding whether the system will be single-blind, double-blind, or onymous (non-anonymous), and determining the number of reviewers per assignment.

Due to the growing interest in peer grading and assessment, several systematic literature reviews have been conducted. For example, Gamage et al.\cite{Gamage03042021} provided a broad classification of peer assessment approaches, including assignment submission methods, reviewer types, assessment approaches, and grading algorithms. Ravikiran\cite{ravikiran2020} offered a systematic review of automatic grading tools and methods for detecting superficial or rough reviews. However, none of these studies focus on a key element of peer grading: review distribution strategies.

The aspect of anonymity was addressed by Panadero et al.~\cite{Panadero17112019} in 2019, who conducted an empirical review on the effects of anonymity in peer assessment, examining 14 studies. They found that anonymity encouraged more critical feedback and reduced social pressure, but it can negatively affect accountability, trust, and interpersonal engagement. However, the other two components of assignment distribution---reviewer assignment strategies and the number of reviewers---have received little attention.

To address this gap, our study examines how different reviewer assignment strategies and varying numbers of reviewers affect the accuracy and fairness of peer grading. It also explores how these factors impact students’ learning outcomes and engagement. We aim to answer the following Research Questions (RQs):
\begin{itemize}
     \item \textbf{RQ1:} What different reviewer assignment strategies are used in peer grading?
     \item \textbf{RQ2:} How does assignment influence the accuracy and fairness of peer grading systems and the learning outcomes of peers?
     \item \textbf{RQ3}: To what extent does the number of peer reviewers affect grading accuracy, reliability, reviewer workload, and learning?
\end{itemize}

\section{Materials and Methods}
We followed the PRISMA framework \cite{PAGE2021790} to ensure the rigor and clarity of our study. The process began with predefined selection criteria focused on peer grading (peer marking). A comprehensive search was conducted across scholarly databases using targeted keywords. After removing duplicates, studies were initially screened by title and abstract, then by full text. The final selection was analyzed to categorize reviewer distribution strategies and assessment cardinality's effect.

\subsection{Source of Information and Eligibility Criteria}
\label{source_of_information}
The data for this review were gathered from a range of reputable academic databases selected for their comprehensive coverage of scholarly literature across multiple disciplines. These sources include ACM Digital Library, Web of Science, EBSCO, ERIC, PsycINFO, IEEE Xplore, and PubMed. Each database was chosen for its focus on high-quality and relevant research to the subject areas addressed in this review. 

Included were publications indexed in one of the seven databases, focusing specifically on peer-reviewed journal articles, conference proceedings, book chapters, and technical articles.  To capture a comprehensive range of approaches within the field of peer grading,  works published between 2010 and 2024 were included in the review. Duplicate studies, books, and retracted papers were discarded. The inclusion and exclusion criteria are given in Table ~\ref{tab:inclusion_exclusion}.

\begin{table}[h!]
\centering
\captionsetup{belowskip=-25pt} 

\renewcommand{\arraystretch}{1.2}
\begin{tabular}{C{0.62\linewidth}@{\hspace{1.5em}}C{0.30\linewidth}}
\toprule
\textbf{Inclusion Criteria} & \textbf{Exclusion Criteria} \\
\midrule
\begin{itemize}[left=0pt,
                itemsep=0pt,
                topsep=0pt,       
                partopsep=0pt,
                parsep=0pt]
    \item Publications indexed in one of the seven databases
    \item Scientific articles, conference proceedings, PhD dissertations, book chapters, and technical reports.
    \item Studies published between 2010 and 2024.
    \item Studies published in the English language.
\end{itemize}
&
\begin{itemize}[left=0pt,
                itemsep=0pt,
                topsep=0pt,       
                partopsep=0pt,
                parsep=0pt]
    \item Duplicate studies.
    \item Books.
    \item Studies that do not address at least one RQ.
    \item Studies with retraction notices or errata.
\end{itemize}
\\
\bottomrule
\end{tabular}
\captionsetup{font=small, labelfont=sc, justification=centering}
\caption{Inclusion and Exclusion Criteria for the Systematic Literature Review.}
\label{tab:inclusion_exclusion}
\end{table}

\subsection{Search Strategy and Study Selection}
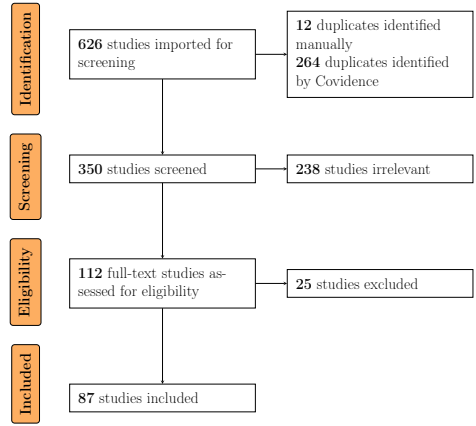
\begin{figure}[htbp]
    \centering
    \captionsetup{belowskip=-10pt} 
    \scalebox{0.28}{\definecolor{myorange}{HTML}{FDAE61}
\tikzset{
    mynode/.style={
        draw, rectangle, align=left, text width=8cm, font=\huge, inner sep=3ex},
    mylabel/.style={
        draw, rectangle, align=center, rounded corners, font=\huge\bf, inner sep=3ex, 
        fill=myorange, minimum height=2.8cm},
    arrow/.style={
        very thick,->,>=stealth}
}

\begin{tikzpicture}[
    node distance=3.6cm,
    start chain=1 going below,
    every join/.style=arrow,
    ]

    \coordinate[on chain=1] (tc);

    \node[mynode, on chain=1, yshift=5cm] (n1)
    {\textbf{626} studies imported for screening\\};

    \node[mynode, join, on chain=1, ] (n2)
    {\textbf{350} studies screened};

    \node[mynode, join, on chain=1] (n3)
    {\textbf{112} full-text studies assessed for eligibility};

    \node[mynode, join, on chain=1] (n4)
    {\textbf{87} studies included};

    \begin{scope}[start chain=going right, node distance=1.5cm]
        \chainin (n1);
        \node[mynode, join, on chain, text width=8cm]
        {\textbf{12} duplicates identified manually\\
        \textbf{264} duplicates identified by  Covidence};

        \chainin (n2);
        \node[mynode, join, on chain, text width=8cm]
        {\textbf{238} studies irrelevant};

        \chainin (n3);
        \node[mynode, join, on chain, text width=8cm]
        {\textbf{25} studies excluded\\
        };
    \end{scope}

    \begin{scope}[start chain=going below, xshift=-6.5cm, node distance=.9cm]
        \node[mylabel, on chain] {\rotatebox{90}{Identification}};
        \node[mylabel, on chain] {\rotatebox{90}{Screening}};
         \node[mylabel, on chain] {\rotatebox{90}{Eligibility}};
        \node[mylabel, on chain] {\rotatebox{90}{Included}};
    \end{scope}

\end{tikzpicture}} 
    \caption{PRISMA flow diagram of the study selection process.}
    \label{fig:prisma}
\end{figure}

An initial search was conducted in the selected databases using the keywords ``peer grading'' and/or ``peer marking'' along with the predefined inclusion and exclusion criteria. This search returned a total of 626 publications. Of these, 276 were identified as duplicates, with 264 detected automatically and an additional 12 flagged manually by the authors. After removing duplicates, 350 papers remained and were screened by abstract to assess their relevance. Following this initial screening, 112 studies were selected for a full-text review. Nine papers were later excluded as they did not address any of the formulated research questions, resulting in a final selection of 87 studies for inclusion in the review (Figure~\ref{fig:prisma}). Of these, 45 were conducted in traditional classroom settings and 42 in MOOCs.

\subsection{Quality Control of Study}
All the authors were actively engaged throughout the research process to ensure the accuracy and trustworthiness of the studies included. At least two authors independently screened and/or reviewed each paper. Any disagreement was followed by discussions to achieve consensus and reduce the likelihood of individual or methodological bias influencing the process. In the screening phase, we achieved a moderate inter-rater agreement ($\kappa$ = 0.55). For the final selection using full-text review, we achieved nearly perfect agreement ($\kappa$ = 0.83)\cite{Landis:Koch:Kappa:Range}. Following paper selection, two authors independently analyzed each paper's reviewer assignment strategies and reviewer quantity. Discussions continued until a consensus was gained.

\section{Results}

This section analyzes our research questions quantitatively and qualitatively.

\subsection{RQ1: Reviewer Assignment Strategies}
We identified four common patterns of assigning reviewers to peer authors: (1) Random Assignment, (2) Competency-Based Assignment, (3) Social Network-Based Assignment, and (4) Bidding-Based Assignment.

\vspace{5pt}

\textbf{Random Reviewer Assignment:} In random assignment, peer review tasks are assigned randomly, without considering any specific reviewer attributes or qualifications \cite{Khatoon2022, Sciarrone2019}. While reviewer selection is random, certain approaches can be incorporated to enhance fairness and grading accuracy. For instance, Chakraborty et al. \cite{Chakraborty2024} assigned each student a total of $k$ submissions to review, with $k/2$ of these being calibration assignments previously graded by the instructor. Random assignment is often used to implement a uniform workload by assigning an equal number of reviews per student ~\cite{LuY2015, García-Martínez2019, Vogelsang2015}. 

\vspace{2pt}

\textbf{Competency-Based Reviewer Assignment:} Reviewers can be allocated based on indicators of reviewer expertise, such as academic performance \cite{H.Lynda2017}, domain knowledge \cite{Bhavya2021}, past grading behavior \cite{Wright2015, Gamage2018, Capuano2016}, and calibration accuracy \cite{Boudria2018}. For example, Bhavya et al. \cite{Bhavya2021} assigned review tasks based on topic similarity, ensuring that reviewers were familiar with the content domain of the projects they evaluated. Similarly, Haddidi et al. \cite{H.Lynda2017} grouped students into clusters based on cognitive indicators such as prior performance, certifications earned, and learning preferences and assigned reviewers accordingly. Sometimes, competency is enforced through strict thresholds. For example, Boudria et al. \cite{Boudria2018} required reviewer teams to pass a calibration threshold ($ \geq 60\%$ agreement with instructor scores) to unlock access to the peer-review phase.

\vspace{2pt}

\textbf{Social Network-Based Reviewer Assignment:} Reviewer assignment can be guided by students’ social connections or prior interactions  \cite{Chiou2015, Jingjing2015}.  Chiou et al. \cite{Chiou2015} used forum-based metrics such as replies, votes, and follows as indicators of social proximity to assign reviewers with low interaction history. Jingjing et. al \cite{Jingjing2015} ensured students did not review classmates, using class membership as a proxy indicator for social closeness.

\vspace{2pt}

\textbf{Bidding-Based Reviewer Assignment:} With bidding, students select the assignments they wish to review rather than being assigned by the instructor. Alipour et al. \cite{Alipour2022} implemented a bidding approach where students could freely choose which assignments to grade. Submissions were anonymous and open to assessment by any participant. Each assignment required a minimum of three reviews, after which reward points decreased to discourage redundant grading.

\begin{figure}[htbp]
    \centering
    \captionsetup{belowskip=-10pt} 
    \scalebox{.6}{\begin{tikzpicture}
    \begin{axis}[
        ybar,
        bar width=3.5mm,
        width=10cm,
        height=4.9cm,
        enlarge x limits=0.35,
        ylabel={Count},
        ymax=60,
        ytick={0,10,...,60},
        symbolic x coords={Traditional Classroom, MOOC, Total},
        xtick=data,
        x tick label style={rotate=0, anchor=center, yshift=-5pt},
        ymin=0,
        legend style={at={(1.05,0.5)}, anchor=west, legend columns=1, /tikz/every even column/.append style={column sep=-0ex}},
        every axis/.append style={font=\small},
        legend cell align={left},
        nodes near coords,
        every node near coord/.append style={font=\footnotesize},
        ymajorgrids=true,
        grid style=dashed,
        legend style={font=\scriptsize, draw=none},
        legend image code/.code={
                \draw[#1, fill=#1, draw=none] (0cm,-0.1cm) rectangle (0.1cm,0.1cm);
            },
        ]

        \addplot+[ybar, fill=blue, text=black] plot coordinates {(Traditional Classroom, 22) (MOOC, 31) (Total, 53)};
        \addplot+[ybar, fill=green, draw=green,
         nodes near coords style={font=\footnotesize,text=black}] plot coordinates {(Traditional Classroom, 8) (MOOC, 7) (Total, 15)};
        \addplot+[ybar, fill=magenta, draw=red, text=black] plot coordinates {(Traditional Classroom, 0) (MOOC, 2) (Total, 2)};
        \addplot+[ybar, fill=red, draw=red, text=black] plot coordinates {(Traditional Classroom, 1) (MOOC, 0) (Total, 1)};

        \legend{Random Reviewer Assignment, Competency-Based Reviewer Assignment, Social Network-Based Reviewer Assignment, Bidding-Based Reviewer Assignment}
    \end{axis}
\end{tikzpicture}} 
    \caption{Distribution of Reviewer Assignment Strategies across literature }
    \label{fig:ra_assignment}
\end{figure}
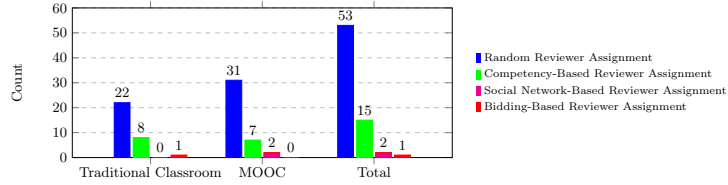

As shown in Figure~\ref{fig:ra_assignment}, random assignment emerges as the most commonly used method, adopted by approximately 75\%  of the systems, followed by competency-based assignment ($\approx 21\%$  of systems). In contrast, social network and bidding-based assignments have received minimal attention, with only one or two systems employing them. For the other 16 papers, there was no explicit mention of reviewer assignment strategies, like whether it is random or any criteria-based.

Random assignment and competency-based matching dominate both in the traditional classrooms and MOOCs. However, competency-based matching is proportionally more prevalent in classrooms, where instructors can draw on course-specific performance data. Social network-based assignment is observed only in MOOCs, likely because these environments generate extensive digital interaction traces that are largely absent in the classroom. 

\subsection{RQ2: Implications of Reviewer Assignment Strategies}

While random assignment is widely favored for its simplicity and scalability, it has its limitations. As highlighted by James et al. \cite{James2018}, some students tend to overgrade, others may harshly underrate, and a few may grade arbitrarily, leading to potential inaccuracies. Random assignment can engender fairness concerns---particularly when low-quality reviewers are unevenly assigned, disproportionately affecting certain submissions. Toll et al. \cite{Toll2017} noted significant variation in the time and effort students dedicate to grading, with some students expressing that their final scores largely depended on their ``luck'' of receiving conscientious reviewers. Additionally, random assignment can produce uneven review distributions without accounting for reviewer participation or engagement. Albano et al. \cite{Albano2017} reported that three students received no evaluations in their study despite each submission being assigned to three reviewers.

However, some of the problems can be mitigated by incorporating different techniques. The harsh-lenient grading problem can be mitigated by using calibration. Yuan et al. \cite{Yuan2020} reduced grading error by up to 39.13 MSE points, even under random reviewer assignment, by incorporating calibration papers. The problem of reviewers not doing their assigned reviews can be mitigated by using dynamic allocation techniques. Andriamiseza et al. \cite{Andriamiseza2023} used a peer review system that selected a submission at random from among those with the fewest current reviews whenever a student began a new review. A similar strategy was employed by CrowdGrader \cite{deAlfaro2014}, where assignments are drawn from submissions that have received the fewest number of reviews.
In some systems, random assignment is intertwined with priority mechanisms. For example, in Staubitz et al. \cite{Staubitz2016}, students who submitted reviews were given a priority boost, increasing the likelihood that their own submissions would be reviewed. This incentivizes active participation in the peer-review process. 

While some mechanisms can be added to random assignment to mitigate its problems, competency-based reviewer assignment provides a more robust solution. This systematically mitigates skewed grading. 
For example, Haddidi et al.\@\cite{H.Lynda2017} and Maicus et al. \cite{Maicus2021} both grouped students into different clusters based on their competency level. They prioritized prior performance, certifications earned, learning preferences,  forum participation, and peer feedback quality as the matrices of competency. Then, each submission was assigned to multiple reviewers with different levels of competency, ensuring that each submission had diverse reviewers. Similarly, Jiménez-Romero et al.\@\cite{Jimenez-Romero2013} classified students into high-reputation and low-reputation groups and ensured that each submission was reviewed by one student from each group and one from a random selection. Badea et al. \cite{Badea2022} created three skill-based reviewer pools---high, medium, and low---and assigned one reviewer from each to every submission. Capuano et al.\@\cite{Capuano2016} proposed a dynamic assignment model that continually matched top-performing reviewers to students who had previously received low-quality reviews, correcting imbalances in feedback quality over time. Along with accuracy and fairness, assigning diverse reviewers has learning benefits. It helps authors to get feedback from different perspectives and makes the feedback complementary \cite{Maicus2021}. 

Some competency-based assignment systems tend to promote meritocracy over collectivism.  Gamage et al.\@\cite{Gamage2018} introduced a feedback-based reputation mechanism where students who gave high-quality feedback were more likely to be reviewed by top-tier graders in future rounds, incentivizing engagement. Wright et al. \cite{Wright2015} separated students into independent and supervised pools based on prior grading accuracy and calibration. Reviewers in the independent pool reviewed only submissions from similarly competent peers, while those in the supervised pool reviewed within their group. Chan et al. \cite{hpchan2016} implemented a reliability-weighted assignment strategy, assigning more reviews to students with high grading reliability to ensure submissions received quality evaluations.

While the effect of social-network assignment has not been measured statistically as a separate variable, it can be expected to outperform random assignment in fairness and objectivity \cite{Chiou2015}. Moreover, bidding-based assignment can speed up grading since reviewer assignment is effectively done by the peers themselves, and might lead students to engage early to to access the best reviewing options.

\subsection{RQ3: Impact of number of reviewers}
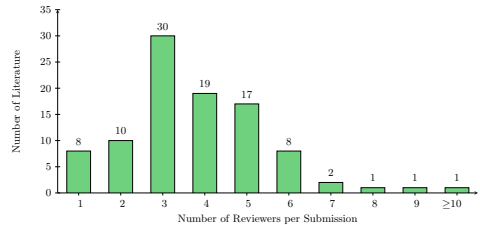
\begin{figure}[htbp]
    \centering
    \captionsetup{belowskip=-10pt} 
    \scalebox{0.45}{\definecolor{myorange}{HTML}{6fd17e}
\begin{tikzpicture}
\begin{axis}[
    ybar,
    bar width=20pt,
    enlarge x limits=false,
    xmin=0, xmax=10,
    ylabel={Number of Literature},
    xlabel={Number of Reviewers per Submission},
    ymin=0, ymax=35,
    ytick={0,5,10,15,20,25,30,35},
    xtick={0.5,1.5,2.5,3.5,4.5,5.5,6.5,7.5,8.5,9.5},
    xticklabels={1, 2, 3, 4, 5, 6, 7, 8, 9, $\geq$10},
    nodes near coords,
    every node near coord/.append style={font=\small, yshift=2pt, text=black},
    xticklabel style={
        font=\small,
        xshift=1pt,
        yshift=-2pt,
        anchor=west,
    },
    axis lines=left,
    width=14cm,
    height=7cm,
    grid=major,
    axis background/.style={fill=white},
    tick style={draw=black},
    major grid style={draw=white!30},
    x tick label style={rotate=1,anchor=east, xshift=6pt,
    yshift=-5pt,},
]
\addplot+[draw=black, fill=myorange] coordinates {
    (0.5,8) (1.5,10) (2.5,30) (3.5,19) (4.5,17)
    (5.5,8) (6.5,2) (7.5,1) (8.5,1) (9.5,1)
};
\end{axis}
\end{tikzpicture}} 
    \caption{Number of Reviewers per Submission across literature }
    \label{fig:per_submission}
    \setlength{\belowcaptionskip}{-10pt}
\end{figure}

We consider two related aspects: (1) the number of reviewers per submission and (2) the number of reviews completed by each student. In most cases, these numbers are identical, except when calibration reviews are assigned or when some students fail to complete their reviews. For simplicity, we treat them as equivalent. Some studies report a range rather than a fixed value, for example, 3--4 per submission \cite{Ou2024}, 4--5 per submission \cite{Zarkoob2024}. For trend analysis, we count each value within a range---a range of 3--5 contributes to counts for 3, 4, and 5. As shown in Figure \ref{fig:per_submission}, the most common configuration is three reviews, followed by four and then five.

A consistent trend across studies is that increasing the number of peer reviews per submission generally enhances the accuracy of grading \cite{Caragiannis2020, Caragiannis2014, Zarkoob2023}. Jimenez-Romero et al. \cite{Jimenez-Romero2013} evaluated accuracy in numeric peer grading and showed a significant error margin when only a few reviewers were used, but observed improved results when more reviews per submission were included. Caragiannis et al. \cite{Caragiannis2014} showed that accuracy improved from  83\% to 93\% when the number of reviews increased from 4 to 10. Similarly, Kulkarni et al. \cite{Kulkarni2013} observed that increasing the number of raters from 5 to 11 led to 7.4\% more grades falling within 10\% of the staff grade. In addition to enhancing grading accuracy, increased participation in review processes can significantly contribute to a student's learning outcomes. Rodgers \cite{Rodgers2019} showed that the shift from one to two reviews resulted in improved alignment with expert grading and was accompanied by student feedback affirming the educational value of reviewing.

Student engagement and perceived workload also tie in with the number of reviews that need to be done. Many studies indicated that requiring more than five reviews per student could lead to disengagement or rushed, low-effort grading that could impact learning gains. Mi et al. \cite{Mi2015} and Fang et al.\@\cite{Fang2017} emphasized this trade-off, recommending a cap of five reviews per student to maintain quality without risking burnout. Lee et al.\@\cite{Lee2019} reported that student engagement remained high over several assignments when five peer reviews were required. Vista et al. \cite{Vista2015} recommended four peer reviews as a good balance between workload and accuracy. The students in Price et al.'s study \cite{Price2016} found the workload of three reviews to be reasonable and beneficial. Chan et al. \@\cite{chanhp2017} found that even three reviews per student could foster meaningful participation if trust and credibility mechanisms were in place.

While accuracy tends to improve with the number of reviewers, this relationship is not linear, and achieving high accuracy does not necessarily require a large number of reviews per submission. Capuano et al.\@\cite{Capuano2016} found that RMSE dropped sharply when increasing the number of reviewers from 1 to 4, showed moderate improvement between 4 and 6, and yielded only marginal gains beyond that point. Nakayama et al. \@\cite{last2024} reported a similar pattern: a substantial accuracy improvement up to 5 reviewers, moderate gains from 5 to 8, slight gains between 8 and 19, and little to no improvement thereafter. Likewise, Raman et al. \cite{Raman2014} showed that while accuracy continued to improve up to around fifteen reviewers, the benefits plateaued beyond that.

Moreover, many studies showed that even three to five reviews could yield high accuracy when applying mechanisms like weighting, calibration, or ordinal models. For instance, Price et al.\@\cite{Price2016} found peer scores with three reviewers to be statistically equivalent to expert scores, with a $R^2$ exceeding 0.8. Fang et al. \cite{Fang2017} introduced a hybrid model incorporating some TA-graded scores to adjust peer-given grades, showing that five reviewers per submission provided reliable results without overwhelming students. James et al. \cite{James2017} showed that five peer reviewers per submission can achieve high accuracy, with a mean absolute error as low as 0.051 for both for a method they call ``Individual Weighting Factor'' and the arithmetic-mean method. However, grading validity suffers considerably when less than two or three reviewers are assigned to each submission ~\cite{last2024}.

\section{Discussion}

This section reflects on key insights from our findings and highlights areas for future research.
\vspace{5pt}

\noindent\textit{Cure vs Prevention:} Many techniques like bias mitigation and noise reduction have been developed for peer-grading systems, primarily to adjust used to adjust already assigned grades. However, reviewer assignment strategies leveraging competency and social networks can serve as proactive measures that reduce grading noise and enhance learning experiences without complex post-processing.

\vspace{2pt}

\noindent \textit{Accuracy vs. Workload Trade-off:} Significant accuracy can be achieved by 3--5 reviews per submission while balancing student learning outcome, engagement, and workload. Increasing the number of peer reviews per submission enhances accuracy; this benefit rapidly diminishes beyond 5--6 reviewers. Systems that rely heavily on large numbers of peer evaluations may risk overburdening students, leading to disengagement or superficial grading. 

\vspace{2pt} 
\noindent \textit{Gaps and Future Directions:} Empirical comparisons between random assignment and competency-based systems remain limited. Large-scale studies that isolate the effects of reviewer competency matching could offer concrete guidance for platform designers. Second, while social-network and bidding approaches have conceptual appeal, they require rigorous testing to assess their impacts on grading objectivity and student trust.  Some approaches combine competency-based assignment with dynamic allocation\cite{reviewer_suggestion}, but little research exists on hybrid distribution models that integrate different reviewer allocation strategies, such as bidding with competency thresholds.

\vspace{2pt}
\noindent \textit{Limitations and threats to validity:} The authors may have overlooked valuable insights from literature and studies not indexed in selected databases. Further, while comprehensive,  the study potentially overlooks relevant non-English studies and gray literature. Finally, the context-specific nature of many studies---such as assignment type and length---may limit the generalizability of our conclusions.
\section{Conclusion}
This systematic review presents different aspects of different reviewer assignment strategies and reviews quantity in optimizing peer grading systems. Our analysis found that, while random assignment remains the most prevalent, competency-based approaches offer greater accuracy and fairness. Three to five reviews per submission strikes an effective balance between grading reliability and student workload. Despite optimistic designs like social networks and bidding-based assignments, further empirical evaluation is needed.  We hope this work serves as a foundation for designing more robust, transparent, and pedagogically meaningful peer grading systems.

\bibliographystyle{IEEEtran}
\bibliography{references} 
\end{document}